\documentclass[a4paper,11pt]{article}

\usepackage{amsmath}

\usepackage{pstricks}

\usepackage{pst-node}

\usepackage[ansinew]{inputenc}

\usepackage{amssymb,amsmath}

\usepackage{amsfonts}

\usepackage{epsfig}

\newtheorem{theorem}{Theorem}

\def\px{\partial_x}

\def\py{\partial_y}

\def\o{\omega}

\def\be{\begin{equation}}       \def\ba{\begin{array}}

\def\ee{\end{equation}}         \def\ea{\end{array}}

\def\bea {\begin{eqnarray}}      \def\eea {\end{eqnarray}}

\def\bean{\begin{eqnarray*}}    \def\eean{\end{eqnarray*}}

\def\RA {\ \Rightarrow\ }

\def\<{\langle} \def\({\left(}  \def\>{\rangle} \def\){\right)}

\newtheorem{exi}{Example}

\author{E. Kartashova$^1$, S. McCallum$^{2}$, \\\\
$^1$ RISC, J.Kepler University, Linz, Austria\\
$^{2}$ Macquarie University, Sydney, Australia\\
\\  e-mail: lena@risc.uni-linz.ac.at, scott@ics.mq.edu.au }

\title{Quantifier elimination for approximate BK-factorization}

\date{}

\begin{document}


\maketitle 

\section{Introduction}
Factorization of linear partial differential operators (LPDOs)
 is a very well-studied problem and a lot of  pure existence
theorems are known. The only known  constructive factorization
algorithm - Beals-Kartashova (BK) factorization - is presented in
\cite{bk2005}). Its comparison with Hensel descent which is
sometimes regarded as constructive, is given in \cite{kr2006}, where
the idea to use BK-factorization for {\it approximate} factorization
is also discussed. It originates in one of the most interesting
features of BK-factorization: at the beginning all the first-order
factors are constructed and afterwards the factorization
condition(s) should be checked. This leads to the important
application area - namely, numerical simulations which could be
simplified substantially if instead of computation with one LPDE of
order $n$ we will be able to proceed computations with $n$ LPDEs all
of order 1. In numerical simulations it is not necessary to fulfill
factorization conditions exactly but with some given accuracy, which
we call approximate factorization.

The idea of the present paper is to look into the feasibility of
solving problems of this kind using {\em quantifier elinination by
cylindrical algebraic decomposition} \cite{Collins:75}. In this
paper we are going to apply this approach to a hyperbolic LPDO of
order 2 with polynomial coefficients.

\section{ Hyperbolic LPDO of order 2}
 A bivariate operator of second order has general form
\begin{equation}\label{A2}
A_2 =a_{20}\px^2+a_{11}\px\py+a_{02}\py^2+a_{10}\px+a_{01}\py+a_{00}
\end{equation}
and is factorizable \cite{bk2005} {\bf iff}
\begin{eqnarray}\label{cond} a_{00}= \mathcal{L} \left\{
 \frac{\o a_{10}+a_{01} - \mathcal{L}(2a_{20} \o+a_{11})}
{2a_{20}\o+a_{11}}\right\}+ \frac{\o a_{10}+a_{01} -
\mathcal{L}(2a_{20} \o+a_{11})}
{2a_{20}\o+a_{11}}\times\nonumber\\
\times \frac{ a_{20}(a_{01}-\mathcal{L}(a_{20}\o+a_{11}))+
(a_{20}\o+a_{11})(a_{10}-\mathcal{L}a_{20})}{2a_{20}\o+a_{11}}.
 \end{eqnarray}
Here  coefficients $a_{i,j}=a_{i,j}(x,y)$ are functions on two
variables $x$ and $y$;  $\o$ is  a distinct root of the following
polynomial $ \mathcal{P}_2(z):=  a_{20}z^2 +a_{11}z +a_{02},\ \
\mathcal{P}_2(\o) = 0$; and $\mathcal{L}$ is a linear differential
operator of the form $ \mathcal{L}=\px-\o\py. $

Let us introduce a function  of two variables $x,y$
\begin{eqnarray} \mathcal{R}= \mathcal{L} \left\{
 \frac{\o a_{10}+a_{01} - \mathcal{L}(2a_{20} \o+a_{11})}
{2a_{20}\o+a_{11}}\right\}+ \frac{\o a_{10}+a_{01} -\mathcal{L}(2a_{20} \o+a_{11})}
{2a_{20}\o+a_{11}}\times\nonumber\\
\times\frac{ a_{20}(a_{01}-\mathcal{L}(a_{20}\o+a_{11}))+
(a_{20}\o+a_{11})(a_{10}-\mathcal{L}a_{20})}{2a_{20}\o+a_{11}},\nonumber
 \end{eqnarray}
and rewrite factorization condition (\ref{cond}) as
$
a_{00}=\mathcal{R}.
$

Now suppose that (\ref{A2}) is a hyperbolic operator in canonical
form, i.e.
\begin{equation}\label{H2}
H_2 =\px^2-\py^2+a_{10}\px+a_{01}\py+a_{00}
\end{equation}
which corresponds to $ a_{20}(x,y)=1, \ a_{11}(x,y)=0, \
a_{02}(x,y)=-1. $ In this case we have two roots $\o_1= 1$ and
$\o_2= -1$, and
 function
$\mathcal{R}$ takes a form
\begin{eqnarray} \label{condH2}\mathcal{R}= \mathcal{L} \left\{
 \frac{a_{10}\pm a_{01}}{2}\right\}+ \frac{(a_{10}\pm a_{01})^2} {4},\nonumber
 \end{eqnarray}
where "+" corresponds to $\o_1= 1$ and "$-$" corresponds to $\o_2=
-1.$ We rewrite (\ref{condH2}) is a slightly different form which
 will more convenient for further use:
\begin{eqnarray} \label{condS}\mathcal{R}= \mathcal{L} \left\{
 S\right\}+ S^2 \quad \mbox{with} \quad S=
\begin{cases}
( a_{10}+a_{01})/2, \ \o=1;\\
 ( a_{10}-a_{01})/2, \ \o=-1.
\end{cases}
 \end{eqnarray}

\section{Polynomial coefficients }
Let us suppose that operator $H_2$ has polynomial coefficients
$a_{ij}$ and regard cases.
\subsection{Polynomials of first degree}
We have 3 polynomials $a_{ij}$ of first degree with two variables
$x,y$: $ a_{00}(x,y)=b_3x+b_2y+b_1, \ a_{10}(x,y)=c_3x+c_2y+c_1, \
a_{01}(x,y)=d_3x+d_2y+d_1,$ then

(1) For  the first root $\o_1= 1$ we have $\mathcal{L} = \px-\py$
and
$$
a_{10}+a_{01}=(c_3+d_3)x+(c_2+d_2)y+(c_1+d_1)= f_3x+f_2y+f_1
$$
with $f_1=(c_1+d_1), \ \ f_2=(c_2+d_2), \ \ f_3=(c_3+d_3). $ Then
$\mathcal{L}( a_{10}+a_{01})=f_3-f_2$ and
\begin{eqnarray}\label{R11} \mathcal{R}_1^{(1)}=
 \frac{ f_3-f_2}{2}+ \frac{( f_3x+f_2y+f_1)^2} {4}
 \end{eqnarray}

(2) For the second root $\o_2= -1$ we have $\mathcal{L} = \px+\py$
and
$$
a_{10}-a_{01}=(c_3-d_3)x+(c_2-d_2)y+(c_1-d_1)= h_3x+h_2y+h_1
$$
with $h_1=(c_1-d_1), \ \ h_2=(c_2-d_2), \ \ h_3=(c_3-d_3). $ Then
$\mathcal{L}( a_{10}-a_{01})=h_3-h_2$ and
\begin{equation}\label{R12} \mathcal{R}_2^{(1)}= \frac{ h_3-h_2}{2}+ \frac{( h_3x+h_2y+h_1)^2}{4}.
 \end{equation}
{\bf Remark 1.} Notice that functions $\mathcal{R}_1$ and
 $\mathcal{R}_2$ coincide symbolically:
\be \label{R1}\mathcal{R}^{(1)}=
 \frac{ s_3-s_2}{2}+ \frac{( s_3x+s_2y+s_1)^2} {4}
\ee
 but of course, the form of $s_i$ as functions of coefficients $c_j, \ \ b_j, \ \
 d_j$ will be different.\\
 {\bf Remark 2.} Factorization condition for the operator
(\ref{H2}) has now very simple form
  $$
\mathcal{R}^{(1)}=a_{00}(x,y) \ \ \RA \ \ \frac{ s_3-s_2}{2}+
\frac{( s_3x+s_2y+s_1)^2} {4}=b_3x+b_2y+b_1
  $$
which yields
$$
\begin{cases}
s_3=0, \ \ s_2s_3=0, \ \ s_2=0, \ \ s_3s_1=2b_3,\\
s_2s_1=2b_2, \ \ s_1^2+2(s_3-s_2)=4b_1
\end{cases}
$$
For instance, in case of the second root this system of equations
has form
$$
\begin{cases}
(c_3-d_3)^2=0, \ \ (c_3-d_3)(c_2-d_2)=0, \ \ (c_2-d_2)^2=0, \\
(c_3-d_3)(c_1-d_1)=2b_3, \ \ (c_2-d_2)(c_1-d_1)=2b_2, \\
(c_1-d_1)^2+(2(c_3-d_3)-(c_2-d_2))=4b_1
\end{cases}
$$
and its solution gives {\bf all exactly factorizable operators} of
this type:
$$
\begin{cases}
a_{00}(x,y)=(c_1-d_1)^2/4\\
a_{10}(x,y)=c_3x+c_2y+c_1\\
a_{01}(x,y)=c_3x+c_2y+d_1
\end{cases}
$$
\subsection{Polynomials of second degree}
Now we have 3 polynomials $a_{ij}$ of second degree  with two
variables $x,y$:
$$
a_{00}(x,y)=b_6x^2+b_5xy+b_4y^2+b_3x+b_2y+b_1,$$$$
a_{10}(x,y)=c_6x^2+c_5xy+c_4y^2+c_3x+c_2y+c_1,$$$$
a_{01}(x,y)=d_6x^2+d_5xy+d_4y^2+d_3x+d_2y+d_1,$$ then

1. For  the first root $\o_1= 1$ we have $\mathcal{L} = \px-\py$ and
$$
a_{10}+a_{01}=(c_6+d_6)x^2+(c_5+d_5)xy+(c_4+d_4)y^2+(c_3+d_3)x+(c_2+d_2)y+(c_1+d_1)$$$$=
f_6x^2+f_5xy+f_4y^2+f_3x+f_2y+f_1
$$
with $f_i=(c_i+d_i), \ \ \forall i=1,2,...,6.$ Then $$\mathcal{L}(
a_{10}+a_{01})=2(f_6x-f_4y)+f_5(y-x)+f_3-f_2$$ and
\begin{eqnarray} \mathcal{R}_1^{(2)}=
 \frac{ 2(f_6x-f_4y)+f_5(y-x)+f_3-f_2}{2}+\nonumber \\
 +\frac{( f_6x^2+f_5xy+f_4y^2+f_3x+f_2y+f_1)^2} {4}
 \end{eqnarray}

2. For the second root $\o_2= -1$ we have $\mathcal{L} = \px+\py$
and
$$
a_{10}-a_{01}=(c_6-d_6)x^2+(c_5-d_5)xy+(c_4-d_4)y^2+(c_3-d_3)x+(c_2-d_2)y+(c_1-d_1)$$$$=
h_6x^2+h_5xy+h_4y^2+h_3x+h_2y+h_1
$$
with $h_i=(c_i-d_i), \ \ \forall i=1,2,...,6.$ Then $$\mathcal{L}(
a_{10}-a_{01})=2(h_6x-h_4y)+h_5(y-x)+h_3-h_2$$ and
\begin{eqnarray} \mathcal{R}_2^{(2)}=
 \frac{ 2(h_6x-h_4y)+h_5(y-x)+h_3-h_2}{2}+\nonumber \\
 +\frac{( h_6x^2+h_5xy+h_4y^2+h_3x+h_2y+h_1)^2} {4}.
 \end{eqnarray}
As above we have in fact one function \bea
\label{R2}\mathcal{R}^{(2)}=
 \frac{ 2(s_6x-s_4y)+s_5(y-x)+s_3-s_2}{2}+\nonumber \\
 +\frac{( s_6x^2+s_5xy+s_4y^2+s_3x+s_2y+s_1)^2} {4}.
\eea
 {\bf Remark 3.} As direct corollaries of linear differentiation
of polynomials one can conclude that for a polynomial any finite
degree $n$, function $ \mathcal{R}^{(n)}$ has form
$$ \mathcal{R}^{(n)}=\sum_{k=1}^{n-1}
\mathcal{R}^{(k)}+(\mbox{other terms})$$ and is a polynomial of
degree $r$ with $r \le 2n$, with necessary condition of exact
factorization being $\ \deg (a_{00}) \le \deg(\mathcal{R}^{(n)})$.

\section{Problem setting for QE}
We use standard formal language of elementary real algebra, that is,
Tarski algebra \cite{Collins:75} and
 formulate a first simple case of approximate
factorization of LPDO as a quantifier elimination (QE) problem.
Namely, we shall consider the case of a hyperbolic LPDO of order 2
with polynomial coefficients of the first order. In this case we
have

(1) three polynomials $a_{ij}(x,y)$ of first degree in the two
variables $x,y$; the coefficients of these polynomials are also
variables:
$$
a_{00}(x,y)=b_3x+b_2y+b_1$$$$ a_{10}(x,y)=c_3x+c_2y+c_1$$$$
a_{01}(x,y)=d_3x+d_2y+d_1;$$

(2) one function $$ \mathcal{R}^{(1)}(x,y)=
 \frac{ s_3-s_2}{2}+ \frac{( s_3x+s_2y+s_1)^2} {4}
 $$ with $s_i$ given by (\ref{R11}) or by (\ref{R12}), i.e. $s_i=c_i+d_i$ for the first root and
 $s_i=c_i-d_i$ for the second root;

(3) a constant $\varepsilon$;

(4) constants $m$ and $n$, which define a bounded rectangular region
in the plane: $-m < x < m$, $-n < y < n$.

{\bf Remark 4.} Notice that the special form of the factorization
condition allowed us to reduce the number of variables needed for
this QE problem. Initially we had 9 variables $b_3, b_2, b_1, c_3,
c_2, c_1, d_3, d_2, d_1 $, but in fact it is enough to consider only
the 6 variables $s_1, s_2, s_3, b_1, b_2, b_3$.\\

With all this given, let us consider the quantified formula of
elementary real algebra $\phi^* = \phi^*(b_i, s_j)$ which asserts
that ``for all $x$ and $y$ in the bounded region $-m < x < m$, $-n <
y < n$, we have $-\varepsilon < a_{00}(x,y) - \mathcal{R}^{(1)}(x,y)
< \varepsilon$.'' We wish to eliminate the quantifiers from
$\phi^*(b_i, s_j)$. More precisely, we wish to find a formula of
elementary real algebra $\phi' = \phi'(b_i, s_j)$, free of
quantifiers, such that if $\phi'(b_i, s_j)$ is true then
$\phi^*(b_i, s_j)$ is true. That is, we wish to find conditions on
the coefficients of the initial polynomials $a_{ij}(x,y)$ which
imply that the function $\mathcal{R}^{(1)}(x,y)$ differs not too
much from one these polynomials, namely $a_{00}(x,y)$, throughout
the bounded region $-m < x < m$, $-n < y < n$.

\section{Synopsis of QE by CAD}

Let $A$ be a set of integral polynomials in $x_1, x_2 \ldots, x_r$,
where $r \ge 1$. An {\em A-invariant cylindrical algebraic
decomposition (CAD)} of ${\bf R}^r$, $r$-dimensional real space, is
a decomposition $D$ of ${\bf R}^r$ into nonempty connected subsets
called cells such that

1. the cells of $D$ are cylindrically arranged with respect to the
variables $x_1, x_2, \ldots, x_r$;

2. every cell of $D$ is a semialgebraic set (that is, a set defined
by means of boolean combinations of polynomial equations and
inequalities); and

3. every polynomial in $A$ is sign-invariant throughout each cell of
$D$.

The CAD algorithm as originally conceived \cite{Collins:75, ACM:84}
has inputs and outputs as follows.
Given such a set $A$ of $r$-variate polynomials and a nonnegative integer
$f$ with $f < r$, the algorithm produces as its output
a description of an $A$-invariant CAD $D$ of ${\bf R}^r$,
in which explicit semialgebraic defining formulas are provided only for
the cells of the CAD $D_f$ of ${\bf R}^f$ induced (that is,
implicitly determined) by $D$.
The description of $D$ comprises lists of indices and sample points
for the cells of $D$. (Every cell is assigned an {\em index}
which indicates its position within the cylindrical structure of $D$.)

The working of the original CAD algorithm can be summarized as
follows. If $r = 1$, an $A$-invariant CAD of ${\bf R}^1$ is
constructed directly, using polynomial real root isolation. If $r >
1$, then the algorithm computes a {\em projection set} $P$ of
$(r-1)$-variate polynomials (in $x_1, \ldots, x_{r-1}$) such that
any $P$-invariant CAD $D'$ of ${\bf R}^{r-1}$ can be extended to a
CAD $D$ of ${\bf R}^r$. If $f = r$ we set $f' \leftarrow f-1$ and
otherwise set $f' \leftarrow f$. Then the algorithm calls itself
recursively on $P$ and $f'$ to get such a $D'$. Finally $D'$ is
extended to $D$. In order to produce semialgebraic defining formulas
for the cells of $D_f$ the algorithm must be used in a mode called
{\em augmented projection}.

Thus for $r > 1$, if we trace the algorithm, we see that it computes
a first projection set $P$, eliminating $x_r$,
then computes the projection of $P$, eliminating $x_{r-1}$, and so on,
until the $(r-1)$-st projection set has been obtained,
which is a set of polynomials in the variable $x_1$ only.
This is called the {\em projection phase} of the algorithm.
The construction of a CAD of ${\bf R}^1$ invariant with respect to
the $(r-1)$-st projection set is called the {\em base phase}.
The successive extensions of the CAD of ${\bf R}^1$ to a CAD of ${\bf R}^2$,
the CAD of ${\bf R}^2$ to a CAD of ${\bf R}^3$, and so on,
until an $A$-invariant cad of ${\bf R}^r$ is obtained,
constitute the {\em extension phase} of the algorithm.

Now we consider the {\em quantifier elimination (QE) problem} for the
elementary theory of the reals: given a quantified formula
(known as a {\em QE problem instance}) of elementary real algebra
\[\phi^* = (Q_{f+1} x_{f+1}) \ldots (Q_r x_r) \phi(x_1, \ldots,x_r)\]
where $\phi$ is a formula involving the variables
$x_1, x_2, \ldots, x_r$ which is free of quantifiers,
find a formula $\phi'(x_1, \ldots, x_f)$,
free of quantifiers, such that $\phi'$ is equivalent to $\phi^*$.
The QE problem can be solved by constructing a certain CAD of ${\bf R}^r$.
The method is described as follows.

1. Extract from $\phi$ the list $A$ of distinct non-zero $r$-variate
polynomials occurring in $\phi$.

2. Construct lists $S$ and $I$ of sample points and cell indices,
respectively, for an $A$-invariant CAD $D$ of ${\bf R}^r$, together
with a list $F$ of semialgebraic defining formulas for the cells of
the CAD $D_f$ of ${\bf R}^f$ induced by $D$.

3. Using $S$, evaluate the truth value of $\phi^*$ in each cell of
$D_f$. (By construction of $D$, the truth value of $\phi^*$ is
constant throughout each cell $c$ of $D_f$, hence can be determined
by evaluating $\phi^*$ at the sample point of $c$.)

4. Construct $\phi'(x_1, \ldots, x_f)$ as the disjunction of the
semialgebraic defining formulas of those cells of $D_f$ for which
the value of $\phi^*$ has been determined to be true.\\

The above algorithm solves any given particular instance of the QE problem
in principle. However the computing time of the algorithm grows steeply
as the number $r$ of variables occurring in the input formula
$\phi$ increases.

Collins and Hong \cite{CollinsHong:91} introduced the method of {\em partial
CAD construction} for QE. This method, named with the acronym
QEPCAD, is based upon the simple observation that we can often solve
a QE problem by means of a partially built CAD. The QEPCAD algorithm
was originally implemented by Hong. A recent implementation, denoted
by QEPCAD-B, contains improvements by Brown, Collins, McCallum,
and others -- see \cite{Brown:03}. QEPCAD-B has solved a range of
reasonably interesting problems for which the original QE algorithm
takes too much time. Nevertheless the worst case computing time of
QEPCAD-B remains large (that is, it depends doubly-exponentially
on $r$).

\section{Application of QEPCAD to BK-factorization}

We consider only the first simple case of approximate factorization
described in Section 4. Using the notation of Section 4,
we suppose that $\varepsilon$,
$m$ and $n$ have been given specific constant values,
say $\varepsilon = m = n = 1$,
and we consider the formula
$\phi^*(b_i, s_j)$ which asserts that
\begin{equation}
  (\forall x)(\forall y)[
  (|x| < 1 \wedge |y| < 1) \Rightarrow
  |a_{00}(x,y) - \mathcal{R}^{(1)}(x,y)| < 1 ].
  \label{phi*}
\end{equation}
We wish to find a formula $\phi'(b_i, s_j)$, free of quantifiers,
such that $\phi'(b_i, s_j)$ implies $\phi^*(b_i, s_j)$.

\vspace{0.5cm} {\bf Remark 5.} It would be of greatest interest to
find the most general such $\phi'(b_i, s_j)$ -- that is, to find
quantifier-free $\phi'(b_i, s_j)$ {\em equivalent} to $\phi^*(b_i,
s_j)$. But as we'll see it seems that the time and space resources
needed to do this are prohibitive. We'll also see that it is not as
time consuming, yet hopefully still of interest, to find
quantifier-free conditions merely {\em sufficient} for $\phi^*$ to
be true.\\

We attempted to find a solution to the above QE problem instance by
running the program QEPCAD-B with the quantified formula \ref{phi*}
(rewritten so that the variables $b_i, s_j$ appear explicitly, and
the denominator 4 is cleared from the right hand side of the
implication, see \ref{phi**} below) as its input. The variable
ordering used was $(s_3, s_2, s_1, b_3, b_2, b_1, x, y)$. The
computer used for this and subsequent experiments was a Sun server
having a 292 MHz ultraSPARC risc processor. Forty megabytes of
memory were made available for list processing. However the program
ran out of memory after approximately one hour and forty minutes.
The program was executing the projection phase of the algorithm when
it stopped. The first three projection steps -- that is, successive
elimination of $y$, $x$ and $b_1$ -- were complete.

Increasing the amount of memory to eighty megabytes did not help --
the program still ran out of memory during the fourth projection
step (that is, during elimination of $b_2$).

\subsection{Searching for quantifier-free sufficient conditions}
Of course a very special, but completely trivial, quantifier-free
sufficient condition for our
QE problem instance is the formula
\[\phi'(b_i, s_j) := [b_1 = 0 \wedge b_2 = 0 \wedge b_3 = 0 \wedge
                    s_1 = 0 \wedge s_2 = 0 \wedge s_3 = 0].\]
It could be of some interest to look for partial solutions to (that
is, quantifier-free sufficient conditions for) our QE problem
instance in which some but not all of the variables $b_i, s_j$ are
equal to zero. For example, recall that the given quantified
(\ref{phi*}) -- after rewriting so that the variables $b_i, s_j$
appear explicitly and the denominator 4 is cleared from the right
hand side of the implication -- is:
\begin{equation}
  (\forall x)(\forall y)[
  (|x| < 1 \wedge |y| < 1) \Rightarrow
  |4 b_3 x + 4 b_2 y + 4 b_1 - 2 (s_3 - s_2) - (s_3 x + s_2 y + s_1)^2| < 4 ].
  \label{phi**}
\end{equation}
Suppose that we put $b_2 = s_2 = 0$ in (\ref{phi**}). We obtain:
\[
  (\forall x)(\forall y)[
  (|x| < 1 \wedge |y| < 1) \Rightarrow
  |4 b_3 x + 4 b_1 - 2 s_3  - (s_3 x + s_1)^2| < 4 ]
\]
which is equivalent to:
\begin{equation}
  (\forall x)[
  (|x| < 1) \Rightarrow
  |4 b_3 x + 4 b_1 - 2 s_3  - (s_3 x + s_1)^2| < 4 ],
  \label{psi*}
\end{equation}
which we shall denote by $\psi^*(b_i, s_j)$.

The following theorem shows that a partial solution
to the special QE problem instance $\psi^*(b_i, s_j)$
(that is, a quantifier-free sufficient
condition for $\psi^*$) leads to a partial solution to
the QE problem instance $\phi^*$ (that is, a quantifier-free
sufficient condition for $\phi^*$).

\begin{theorem}
Suppose that $\psi'(b_i, s_j)$ is a quantifier-free formula,
involving only $b_1, b_3, s_1, s_3$, which implies
$\psi^*(b_i, s_j)$. Then the quantifier-free formula
$\psi'(b_i, s_j) \wedge b_2 = 0 \wedge s_2 = 0$ implies
$\phi^*(b_i, s_j)$.
\end{theorem}

$\blacktriangleright$ Let $b_i, s_j$ be real numbers. Assume
$\psi'(b_i, s_j) \wedge b_2 = 0 \wedge s_2 = 0$. Then $\psi^*(b_i,
s_j) \wedge b_2 = 0 \wedge s_2 = 0$ is true, by hypothesis. Take
real numbers $x$ and $y$, with $|x| < 1$ and $|y| < 1$. Then
\[ |4 b_3 x + 4 b_2 y + 4 b_1 - 2 (s_3 - s_2) - (s_3 x + s_2 y + s_1)^2|
   =
   |4 b_3 x + 4 b_1 - 2 s_3 - (s_3 x + s_1)^2|
   < 4,
\]
by virtue of (\ref{psi*}) (since $|x| < 1$). Hence (\ref{phi**}) is
true. $\blacksquare$\\

The above discussion suggests that it would be worthwhile to try to
find a solution to the simplified, special QE problem instance
$\psi^*$ using the program QEPCAD-B. Putting (\ref{psi*}) into a
slightly more general form, and hence reducing by 1 the number of
variables in the formula, we obtain:
\begin{equation}
(\forall x)[
  (|x| < 1) \Rightarrow
  |a x^2 + b x + c | < 4].
  \label{theta*}
\end{equation}
A partial solution $\theta'(a,b,c)$ to (\ref{theta*}) could easily
be transformed into a partial solution $\psi'(b_1, b_3, s_1, s_3)$
to $\psi^*$ by setting $a = -s_3^2$, $b = 4b_3 - 2 s_1 s_3$ and $c =
4b_1 - 2s_3 - s_1^2$.

We ran program QEPCAD-B with (\ref{theta*}) as its input. Eighty
megabytes of memory were made available for list processing. After
191 seconds the program produced the following quantifier-free
formula equivalent to (\ref{theta*}):
\begin{verbatim}
c - b + a + 4 >= 0 /\ c - b + a - 4 <= 0 /\
c + b + a + 4 >= 0 /\ c + b + a - 4 <= 0 /\
[ 4 a c - b^2 + 16 a > 0 \/ 4 a c - b^2 - 16 a > 0 \/
[ b^2 - 16 a = 0 /\ b^2 + 16 a > 0 ] \/ [ b^2 - 16 a < 0 /\ b - 2 a >= 0 ] \/
[ b^2 - 16 a < 0 /\ b + 2 a <= 0 ] \/ [ b^2 - 16 a > 0 /\ b + 2 a >= 0 ] \/
[ b^2 - 16 a > 0 /\ b - 2 a <= 0 ] \/
[ b^2 - 16 a = 0 /\ c - b + a + 4 > 0 /\ c - b + a - 4 < 0 ] ].
\end{verbatim}
Since $a = -s_3^2$, we have $a \le 0$. We ran QEPCAD-B a second time,
this time using the command
\begin{verbatim}
assume [a <= 0].
\end{verbatim}
After 60 seconds the program produced the following somewhat simpler
quantifier-free formula equivalent to (\ref{theta*}) under the
assumption $a \le 0$:
\begin{verbatim}
c - b + a + 4 >= 0 /\ c - b + a - 4 <= 0 /\
c + b + a + 4 >= 0 /\ c + b + a - 4 <= 0 /\
[ 4 a c - b^2 - 16 a > 0 \/ [ b > 0 /\ b + 2 a >= 0 ] \/
[ b < 0 /\ b - 2 a <= 0 ] \/
[ b^2 + 16 a = 0 /\ c - b + a + 4 > 0 /\ c - b + a - 4 < 0 ] ].
\end{verbatim}
It is possible to induce the program to produce an arguably even simpler
solution formula using less computing time
by making two separate runs of QEPCAD-B. The first run uses the command
\begin{verbatim}
assume [a < 0].
\end{verbatim}
After just 1.9 seconds the program produced the following
quantifier-free formula equivalent to (\ref{theta*}) under the
assumption $a < 0$:
\begin{verbatim}
c - b + a + 4 >= 0 /\ c - b + a - 4 <= 0 /\
c + b + a + 4 >= 0 /\ c + b + a - 4 <= 0 /\
[ b - 2 a <= 0 \/ b + 2 a >= 0 \/ 4 a c - b^2 - 16 a > 0 ]. (15)
\end{verbatim}
The above formula is perhaps the most elegant and understandable of
those obtained by applying QEPCAD-B to Formula \ref{theta*}. For it
is a slight improvement of (that is, slightly more compact than) a
formula seen to be equivalent to it (under assumption $a < 0$) which
is quite straightforward to derive by hand from (\ref{theta*}) using
elementary properties of the parabola $y = ax^2 + bx + c$ on the
interval $(-1,+1)$:
\begin{verbatim}
[ 2 a - b >= 0 /\ a + b + c + 4 >= 0 /\ a - b + c - 4 <= 0] \/
[2 a + b >= 0 /\ a - b + c + 4 >= 0 /\ a + b + c -4 <= 0] \/
[2 a - b < 0 /\ 2 a + b < 0 /\ 4 a c - b^2 - 16 a > 0 /\
 a - b + c + 4 >= 0 /\ a + b + c + 4 >= 0].              (16)
\end{verbatim}

{\bf Remark 6.} To derive  by hand  (16) from (\ref{theta*}) under
the assumption $a < 0$, one has to notice that function $f(x) = a
x^2 + b x + c$ has its maximum value for $f'(x) = 2 a x + b = 0$,
that is, for $x = -b/(2a)$, and consider  three cases separately:
(1)
 $-b/(2a) \le -1$, (2) $-b/(2a) \ge +1$, and (3)
 $-1 < -b/(2a) < +1$.
For each of the above three cases one can then write down necessary
and sufficient conditions for (\ref{theta*}) to be true. For
example, in Case 1, (\ref{theta*}) is clearly equivalent to $-4 \le
a + b + c \wedge a - b + c \le 4$. After treating each of the above
cases, we obtain (16) by forming the disjunction of
the formulas corresponding to the cases. \\

Of course, (15) for $a < 0$ is not quite a complete solution to the
QE problem instance of (\ref{theta*}) under assumption $a \le 0$. To
obtain a complete solution we still needed to run QEPCAD a second
time, this time for the case $a = 0$. For the second run we put $a =
0$ in (\ref{theta*}) and use the command
\begin{verbatim}
assume [b /= 0].
\end{verbatim}
After 60 milliseconds the program produced the following formula
equivalent to (\ref{theta*}) with $a = 0$ under assumption $b \neq
0$:
\begin{verbatim}
c - b + 4 >= 0 /\ c - b - 4 <= 0 /\
c + b + 4 >= 0 /\ c + b - 4 <= 0                       (17)
\end{verbatim}
This is immediately seen to be correct! Finally we could obtain a
complete solution to (\ref{theta*}) for $a \le 0$ by combining (15)
for $a < 0$, (17) for $b \neq a = 0$  and the formula
 $\verb|c - 4 < 0 /\ c + 4 > 0|$ (for $a = b = 0$). In fact a simple and elegant way
to achieve such a combination is to insert the disjunct \verb|a = 0|
into the last conjunct of (15):
\begin{verbatim}
c - b + a + 4 >= 0 /\ c - b + a - 4 <= 0 /\
c + b + a + 4 >= 0 /\ c + b + a - 4 <= 0 /\
[ b - 2 a <= 0 \/ b + 2 a >= 0 \/
4 a c - b^2 - 16 a > 0 \/ a = 0].                      (18)
\end{verbatim}

\section{Discussion}

As we remarked in Section 5 the worst case computing time of
QEPCAD-B grows steeply as the number of variables in the given QE
problem instance increases. Indeed, as is suggested by the results
reported in Section 6, a complete solution of the QE problem
instance (\ref{phi**}) by QEPCAD-B using a reasonable amount of time
and space seems to be unlikely for the foreseeable future.

Nevertheless the results of Section 6 also suggest that QEPCAD-B
could be of help in searching for certain kinds of sufficient
conditions for (\ref{phi*}), especially those which involve setting
some of the variables to zero.

We briefly mention here another kind of approach which a person
could use to derive another kind of sufficient condition for
(\ref{phi*}) by hand. Namely, one could begin by expanding the
polynomial $a_{00}(x,y) - \mathcal{R}^1(x,y)$ in terms of $x$ and
$y$:
\begin{eqnarray*}
a_{00}(x,y) - \mathcal{R}^1(x,y)  =  (-s_3^2/4) x^2 +
                                       (-2s_3 s_2/4)xy + (-s_2^2/4) y^2 + \\
 (b_3 - (s_1 s_3)/2) x + (b_2 - (s_1 s_2)/2) y +
                                       (b_1 - (s_3 - s_2)/2 - s_1^2/4).
\end{eqnarray*}
By inspection of the terms on the right hand side of the above
equation we see that a sufficient condition for (\ref{phi*}) is:
\begin{verbatim}
| s_3^2 / 4 | < 1 / 6 /\ | 2 s_3 s_2 / 4 | < 1 / 6 /\
| s_2^2 / 4 | < 1 / 6 /\ |b_3 - (s_1 s_3) / 2 | < 1 / 6 /\
| b_2 - (s_1 s_2) / 2 | < 1 / 6 /\
| b_1 - (s_3 - s_2) / 2 - s_1^2 / 4 | < 1 / 6.
\end{verbatim}
The above sufficient condition is unlikely to be obtained in a
reasonable amount of time and space using QEPCAD-B applied to
(\ref{phi**}), even if one issues \verb+assume+ commands. The number
of variables involved is probably too big. However a version of
QEPCAD-B which is planned for the future, which will have the
capability to determine adjacency relationships amongst the cells of
the partial CAD, could be of some use in analyzing certain
topological properties of the truth set in six-dimensional space of
the quantifier-free formula in $b_i, s_j$ above.

\section*{Acknowledgements}
This paper has its origin in discussions between the authors at
RISC-Linz during the second of S.M. to RISC-Linz in 2005. S.M. would
like to thank Professors Franz Winkler and Bruno Buchberger, and all
their colleagues and staff at RISC, for their hospitality during his
stay. S.M. would also like to acknowledge helpful discussions and
communications with Professors Daniel Lazard and Chris Brown. E.K.
acknowledges the support of the Austrian Science Foundation (FWF)
under projects SFB F013/F1304.


\begin{thebibliography}{99}

\bibitem{bk2005} R. Beals, E. Kartashova. "Constructively factoring linear partial differential operators in
two variables." {\it  TMPh} {\bf 145}(2): 1510-1523 (2005)

\bibitem{kr2006} E. Kartashova, O. Rudenko. "Invariant Form of BK-factorization and
its Applications." {\it Proc. GIFT-2006}: 225-241. Eds.: J. Calmet,
W.M. Seiler, R.W. Tucker. Universitätsverlag Karlsruhe (2006)

\bibitem{Collins:75} G. E. Collins. ``Quantifier elimination
for real closed fields by cylindrical algebraic decomposition.''
{\bf LNCS 33}: 134--183. Springer Verlag, Berlin (1975)

\bibitem{ACM:84} D. Arnon, G. Collins, S. McCallum.
``Cylindrical algebraic decomposition I: the basic algorithm.'' {\it
JSC}, {\bf 13}(4): 878--889, (1984)

\bibitem{Brown:03} C. Brown. ``QEPCAD B: a program for computing with
semi-algebraic sets using CADs.'' {\it ACM SIGSAM Bulletin}, {\bf
37}(4): 97--108, (2003).

\bibitem{CavJohnson:98} B. F. Caviness, J. R. Johnson (Eds.)
{\bf Quantifier Elimination and Cylindrical Algebraic
Decomposition.} Springer Verlag, Berlin (1998)

\bibitem{CollinsHong:91} G. E. Collins, H. Hong.
``Partial cylindrical algebraic decomposition for quantifier
elimination.'' {\it JSC}, {\bf 12}(3): 299--328, (1991)

\end{thebibliography}
\end{document}